\documentclass[11pt]{article}

\usepackage[margin=1in]{geometry}
\usepackage{graphicx}
\usepackage{dsfont}
\usepackage[pdftex,colorlinks=true,linkcolor=blue,citecolor=blue,urlcolor=black]{hyperref}
\usepackage{amsmath, amsthm, amssymb,amsfonts,bbm,bm}
\usepackage{subfigure}
\usepackage{comment}
\usepackage{soul}
\usepackage{url}
\usepackage{pdflscape}
\usepackage[ruled,lined,linesnumbered]{algorithm2e}
\usepackage[dvipsnames]{xcolor}
\usepackage[colorinlistoftodos]{todonotes}
\usepackage{array}
\usepackage{cite}
\usepackage{authblk}
\usepackage{siunitx}
\usepackage{placeins}

\newcommand{\cN}{\mathcal{N}}

\newcommand{\ket}[1]{|#1 \rangle}
\newcommand{\bra}[1]{\langle#1|}

\newcommand{\id}{\mathbbm{I}}

\newcommand{\tr}{\operatorname{Tr}}

\newcommand{\vecsig}[0]{\ensuremath{\bm{\sigma}}\xspace}

\newcommand{\cnot}{\text{CNOT}}
\newcommand{\D}{\langle D^2\rangle}

\newcommand{\be}{\begin{equation}}
\newcommand{\ee}{\end{equation}}
\newcommand{\bea}{\begin{eqnarray}}
\newcommand{\eea}{\end{eqnarray}}
\newcommand{\bes}{\begin{equation*}}
\newcommand{\ees}{\end{equation*}}
\newcommand{\beas}{\begin{eqnarray*}}
\newcommand{\eeas}{\end{eqnarray*}}

\makeatletter
\newtheorem*{rep@theorem}{\rep@title}
\newcommand{\newreptheorem}[2]{%
\newenvironment{rep#1}[1]{%
 \def\rep@title{#2 \ref{##1} (restated)}%
 \begin{rep@theorem}}%
 {\end{rep@theorem}}}
\makeatother

\newtheorem*{thm*}{Theorem}

\newtheorem*{lem*}{Lemma}

\newreptheorem{thm}{Theorem}

\begin{document}
\title{Minimising statistical errors in calibration of quantum-gate sets}

\author[1]{Yaiza Aragonés-Soria}

\author[2]{René Otten}
 
\author[2]{Tobias Hangleiter}

\author[2]{Pascal Cerfontaine}
 
\author[1]{David Gross}

\affil[1]{Institute for Theoretical Physics, University of Cologne, Zülpicher Str. 77, 50937 Köln, Germany}
\affil[2]{JARA-FIT Institute for Quantum Information, Forschungszentrum
Jülich GmbH and RWTH Aachen University, 52074 Aachen, Germany}

\date{\today}

\maketitle

\begin{abstract}
Calibration of quantum gates is a necessary hurdle to overcome on the way to a reliable quantum computer. In a recent paper, a protocol called Gate Set Calibration protocol (GSC) has been introduced and used to learn coherent errors from multi-qubit quantum gates. 
Here, we extend this study in a number of ways:
First, we perform a statistical analysis of the measurement uncertainties.
Second, we find explicit measurement settings that minimize this uncertainty, while also requiring that the protocol involves only a small number of distinct gates, aiding physical realizability.
We numerically demonstrate that, just by adding two more single-qubit gates to GSC, the statistical error produced in the calibration of a $\cnot$ gate is divided by a factor of more than two.
\end{abstract}

\section{Introduction}
A fundamental property of quantum mechanics is its intrinsic uncertainty: responses of measurements can only be predicted probabilistically. 
Therefore, experiments aimed to extract information from a quantum system have to be designed so as to reduce the expected statistical uncertainty.
The branch of statistics concerned with such questions is called  \emph{design of experiments} (DOE). Among the main objectives of DOE is the optimisation of statistical error, as well as validity, reliability and replicability. DOE has been applied in different quantum information problems such as parameter estimation \cite{DOEapplied1,DOEapplied2}, and quantum state and process tomography \cite{DOEapplied3,DOEapplied4,DOEapplied5}. In this article, we address the task of implementing DOE in a calibration protocol for quantum gate sets.

Specifically, we will work with the Gate Set Calibration (GSC) protocol introduced in Ref.~\cite{GSCpaper} for the purpose of identifying coherent errors in unitary quantum gates.
 
Given a set of available quantum gates, initial states and measurements, GSC constructs short gate sequences that allow all coherent gate errors to be extracted. This protocol can be considered as a middle way between Randomized Benchmarking (RB) and Gate Set Tomography (GST), as it provides more detailed information than the former using fewer measurements than the latter \cite{RB,GST}.

In this article we use DOE to optimise GSC. Specifically, we introduce parameters in the sequences that GSC builds and perform an optimisation over these parameters to minimise the statistical error of GSC. 

This article is structured as follows. Section \ref{s:GSC} introduces a simplified version of GSC and establishes the notation. In Section \ref{s:statistics} we present the theoretical framework under which the optimisation of GSC will be carried out. In other words, we analyse the statistics of GSC and define a measure of uncertainty that will allow us to compare the statistical error of different versions of GSC. Section \ref{s:optimizing} is dedicated to the optimisation of GSC. We design an optimisation protocol to minimise the statistical error with respect to the measure of uncertainty. Moreover, the version of GSC resulting from the optimisation is further modified taking into account the feasibility of the protocol. Finally, we verify in Section \ref{s:imperfect} that the reduction of the statistical error is also achieved when imperfect measurements are considered. The paper ends with an outlook in Section \ref{s:outlook}.

\section{The Gate Set Calibration protocol} \label{s:GSC}
In Ref.~\cite{GSCpaper}, a new protocol for characterizing and calibrating quantum gates, \emph{Gate Set Calibration} protocol (GSC), has been introduced. 
In this section, we will introduce the notation and present a slightly simplified version of the protocol. In particular, we restrict attention to the specific, and experimentally relevant, case of calibrating a two-qubit $\cnot$ gate subject to coherent errors.

Gate Set Calibration consists of the following steps. 
First, the qubits are initialized in a known state. 
Then, a gate sequence from a set of available gates is applied to the initial state, and finally a measurement is performed. 
All coherent gate errors can be extracted by repeating this process for different sequences, initial states and measurements, as we will see below.

Assume that we aim to implement a set of unitary quantum gates. In reality, the hardware will realize perturbed gates, which we still assume to be unitary. 

Consider concretely a situation where only gates in the set $\{\cnot,\text{X}^{(1)}_{\theta},\text{Y}^{(1)}_{\theta},\text{X}^{(2)}_{\theta},\text{Y}^{(2)}_{\theta}\}$ can be applied, namely a global  $\cnot$ and arbitrary single-qubit rotations around X- and Y-axis. Here, the superscripts indicate on which qubit the gate is applied, and the gates are defined as
$$
\cnot:=\ket{0}\bra{0}\otimes\id+\ket{1}\bra{1}\otimes\sigma_1,~X_{\theta}:=e^{-i\frac{\theta}{2}\sigma_1},~Y_{\theta}:=e^{-i\frac{\theta}{2}\sigma_2},
$$
where $\sigma_1$, $\sigma_2$, $\sigma_3$ are the Pauli matrices and $\sigma_0=\id$.   
We assume that all single-qubit gates are already calibrated, and thus here we want to use GSC to tune a perturbed $\cnot$ gate, which we denote as $\tilde{\cnot}$. 
The deviation from the optimal gate is measured by the \emph{coherent error operator} associated with the gate. In the particular case of the $\cnot$ gate we define
\begin{equation}\label{e:E}
    E:=\cnot^{-1}\tilde{\cnot}.
\end{equation}
In order to analyze the perturbed gate, we expand the error operator in the Pauli basis
$$
E(\vec{p})=\id-i\sum_{k=1}^{15}p_k\tau_k,
$$
for suitable coefficients $\vec{p}$, which we call \emph{error parameters}, and for $\tau_k=\tau_{ij}=\sigma_i\otimes\sigma_j$.
The choice of the prefactor of the sum ensures that the unitarity of $E$ to first order implies that $p_k$ are real.

The ultimate goal of GSC is to obtain the error parameters, $\vec{p}$. 
To estimate the error parameters, GSC considers a number of experimental settings, which we label with the letter $s$. These settings can vary in the initial state, the gate sequence, and the measurement. As initial state we will always take $\rho=\ket{00}\bra{00}$ and perform a measurement $M_s\in\{\sigma_0\otimes \sigma_3, \sigma_3\otimes \sigma_0\}$.
Thus, for each setting one can experimentally estimate the quantity
\begin{equation}\label{e:Rs}
R_s(\vec p)
=
\tr\left[(G_{l_s,s}\cdots\tilde{\cnot}(\vec{p})\cdots G_{1,s})\ket{00}\bra{00}(G^\dag_{1,s}\cdots\tilde{\cnot}(\vec{p})\cdots G^\dag_{l_s,s})M_s\right],
\end{equation}
which we will refer to as the \emph{measurement responses}. Here, $l_s$ is the length of each sequence and $G_{j,s}\in\{\text{X}^{(1)}_{\theta},\text{Y}^{(1)}_{\theta},\text{X}^{(2)}_{\theta},\text{Y}^{(2)}_{\theta}\}$ are the single-qubit gates of sequence $s$.

Following the framework of Ref.~\cite{GSCpaper}, we now make the assumption that the error parameters are small, and thus assume that the vector of responses depends only linearly on $\vec{p}$. We write
\begin{equation} \label{e:Rsapprox}
\vec{R}(\vec{p})=\vec{R}(\vec{0})+L\vec{p},
\end{equation}
where we define the matrix elements of $L$ as
$$
L_{su}:=\left.\dfrac{\partial R_s(\vec{p})}{\partial p_u}\right|_{\vec{p}=\vec{0}}.
$$
These matrix elements, $L_{su}$, as well as $\vec{R}(\vec{0})$ can be computed analytically. Therefore, given the perturbed responses, $\vec{R}(\vec{p})$, we obtain the gate errors parameters by solving the inverse problem 

\begin{equation}\label{e:vecp}
\vec{p}=L^{-1}\left[\vec{R}(\vec{p})-\vec{R}(\vec{0})\right]. 
\end{equation}
To ensure that the matrix $L$ is non-singular, it is necessary that the number of responses is equal to or greater than the number of error parameters. When this is not the case, additional responses can be added by considering more sequences, initial states or measurements. Moreover, the condition number of $L$ must not be too large so that computing the inverse of L is numerically stable.

\begin{figure}[htb]
\centering
\includegraphics[scale=0.3]{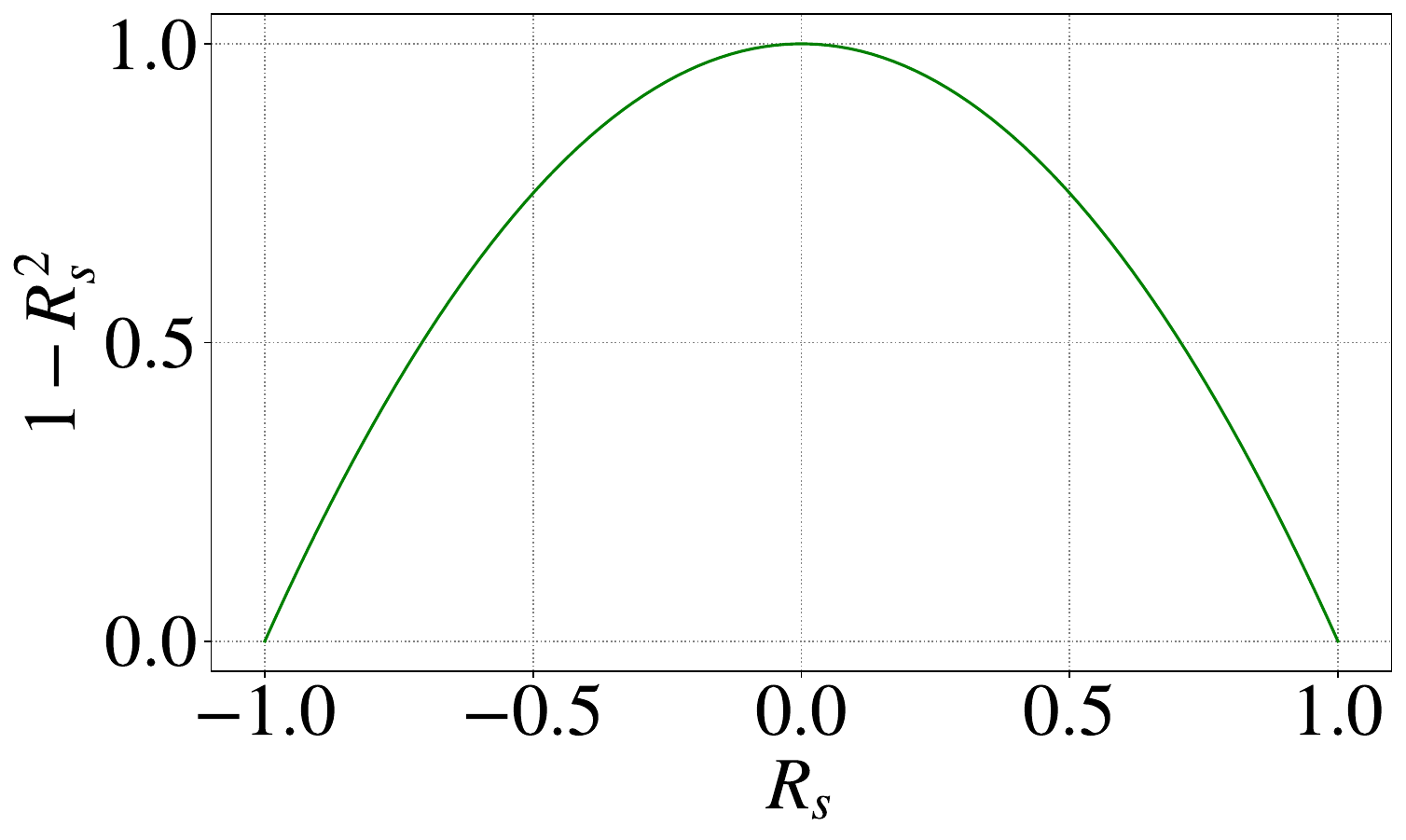}
\caption{Variance of a estimated response, $R^*_s$, times the number of repetitions, $N$, as a function of the expectation value (see Eq.~(\ref{e:Sigma})).}
\label{f:variance}
\end{figure}

Having introduced the basic setup, we are now in a position to explain more clearly the motivation of this paper. Table \ref{t:sequences} summarizes the combination of sequences and measurements used in Ref.~\cite{GSCpaper} to calibrate a perturbed $\cnot$ using GSC. In the third column of Tab.~\ref{t:sequences}, we see that the settings chosen consider only responses distributed around zero. Put differently, $R_s(\vec{0})=0$ for all $s=1,\dots,15$. It turns out that, if one looks at statistics, this is also the point of maximal uncertainty since the variance of the responses is maximal for $R_s(\vec{0})=0$ (see Fig.~\ref{f:variance}). In fact, it is not obvious what the optimal point is, i.e., which settings of GSC minimize the statistical error and, at the same time, ensure stability of the numerical inverse of $L$ and implementability of the protocol. Our research looks into this trade-off between reducing the statistical error, but still having a well-conditioned matrix $L$. Although we only consider the calibration of a two-qubit gate, the statistical analysis performed in the following sections can be generalised to higher dimensions straightforward.

\begin{table}[phtb]
\begin{center}
{\renewcommand{\arraystretch}{2}
\begin{tabular}{|c|c|c|c|c|c|c|}
\hline
$s$ & 1st gate & 2nd gate & 3rd gate & 4th gate & $M_s$ & $R_s(\vec{p})$ \\ 
\hline 
1 & $\cnot$ & $\text{X}^{(1)}_{\pi/2}$ &  &  & $\tau_{12}$ & $0-2 p_5+2 p_{10}$ \\ 
 
2 & $\text{X}^{(1)}_{\pi/2}$ & $\cnot$ &  &   & $\tau_{12}$ & $0-2 p_4-2 p_7$ \\ 
 
3 & $\cnot$ & $\text{Y}^{(1)}_{\pi/2}$ &  &   & $\tau_{12}$ & $0-2 p_6-2 p_9$ \\ 
 
4 & $\text{Y}^{(1)}_{\pi/2}$ & $\cnot$ &   &   & $\tau_{12}$ & $0-2 p_8-2 p_{11}$ \\ 

5 & $\cnot$ & $\text{X}^{(2)}_{\pi/2}$ &   &   & $\tau_3$ & $0-2 p_1-2 p_{13}$ \\ 

6 & $\cnot$ & $\text{Y}^{(2)}_{\pi/2}$ &   &   & $\tau_3$ & $0-2 p_2-2 p_{14}$ \\ 

7 & $\text{X}^{(1)}_{\pi/2}$ & $\cnot$ & $\text{X}^{(1)}_{\pi/2}$ &   & $\tau_{12}$ & $0-2 p_2+2 p_{10}$ \\

8 & $\text{X}^{(1)}_{\pi/2}$ & $\cnot$ & $\text{Y}^{(1)}_{\pi/2}$ &   & $\tau_{12}$ & $0-2 p_6-2 p_{13}$ \\

9 & $\text{Y}^{(1)}_{\pi/2}$ & $\cnot$ & $\text{X}^{(1)}_{\pi/2}$ &   & $\tau_{12}$ & $0+2 p_{10}+2 p_{13}$ \\

10 & $\text{X}^{(1)}_{\pi/2}$ & $\text{X}^{(2)}_{\pi/2}$ & $\cnot$ &   & $\tau_3$ & $0-2 p_7-2 p_{13}$ \\ 

11 & $\text{Y}^{(1)}_{\pi/2}$ & $\cnot$ & $\text{Y}^{(1)}_{\pi/2}$ &   & $\tau_{12}$ & $0-2 p_2-2 p_6$ \\ 

12 & $\text{Y}^{(1)}_{\pi/2}$ & $\text{Y}^{(2)}_{\pi/2}$ & $\cnot$ &   & $\tau_3$ & $0-2 p_{11}-2 p_{14}$ \\ 

13 & $\text{Y}^{(2)}_{\pi/2}$ & $\cnot$ & $\text{X}^{(2)}_{\pi/2}$ &   & $\tau_3$ & $0+2 p_3+2 p_{15}$ \\ 

14 & $\text{X}^{(1)}_{\pi/2}$ & $\cnot$ & $\cnot$ & $\text{Y}^{(1)}_{\pi/2}$ & $\tau_{12}$ & $0-2 p_3-4 p_{12}-2 p_{15}$ \\ 

15 & $\text{X}^{(1)}_{\pi/2}$ & $\text{Y}^{(2)}_{\pi/2}$ & $\cnot$ & $\text{X}^{(2)}_{\pi/2}$ & $\tau_3$ & $0-2 p_6+2 p_{15}$ \\ 
\hline
\end{tabular}}
\caption{List of settings to calibrate a $\tilde{\cnot}$ gate (Eq.~(\ref{e:E})). Each row corresponds to one setting, where gates are applied from left to right and followed by a measurement (either $\tau_3=\sigma_0\otimes\sigma_3$ or $\tau_{12}=\sigma_3\otimes\sigma_0$). The last column contains the measurement responses up to first order in the error parameters, $\vec{p}$ (see Eqns.~(\ref{e:Rs}) and (\ref{e:Rsapprox})).}
\label{t:sequences}
\end{center}
\end{table}

\newpage

\section{The statistics of GSC}\label{s:statistics}
Quantum mechanics is unable to predict with certainty future events, such as the measurement outcomes. Therefore, as we do in this section, it is essential to consider the statistics of the problem we are investigating to minimize statistical errors. Here, we study the statistics of GSC by determining the probability distribution according to which the measurement responses are distributed, as well as the error parameters. Subsequently, we define a quality measure that will allow us to compare the statistical error between different versions of GSC.

Before starting the discussion about GSC statistics, let us establish that all estimated quantities will be denoted by a superscript $*$. Assuming independence, a two-outcome measurement is described by a binomial distribution, $B(N,q)$, where $N$ is the number of repetitions and $q$ the probability of obtaining the outcome. In the scenario presented in Section~\ref{s:GSC}, fifteen measurements are performed, one per setting, and each one is repeated $N$ times. 
Therefore, the overall outcome is described by a random vector with binomially distributed components. 
It is well known that a binomial distribution can be approximated by a normal distribution such that $\cN(Nq, Nq(1-q))$ in the limit of sufficiently large $N$. Our interest is, in particular, focused on spin qubits, which are considered in Ref.~\cite{GSCpaper} and where a high number of repetitions is easy to obtain. We can therefore assume that the overall GSC outcome approximates a multivariate normal distribution and assume specifically that the estimated vector of responses is distributed as $\vec{R}^*(\vec{p})\sim\cN(\vec{R}(\vec{p}),\Sigma)$, where
\begin{equation}\label{e:Sigma}
\Sigma_{su}=\dfrac{1-R^2_s(\vec{p})}{N}\delta_{su},
\end{equation}
with $R_s(\vec{p})$ defined in Eq.~(\ref{e:Rs}). Note that the covariance matrix $\Sigma$ is diagonal, i.e., settings are independent of each other.

The estimated error parameters, $\vec{p}^*$, are computed with Eq.~(\ref{e:vecp}) using the estimated responses, $\vec{R}^*$ obtained from the lab. Therefore, $\vec{p}^*$ is a random vector distributed according to a multivariate Gaussian distribution with $\vec{p}^*\sim\cN(\vec{p},L^{-1}\Sigma L^{-T}),$ where $L^{-T}:=(L^{-1})^T$.

In order to quantify the statistical error, we need to construct a measure of uncertainty. We take our quality measure to be the mean squared distance between $\vec{p}^*$ and $\vec{p}$, i.e.,
\begin{align}\label{e:D2}
\D &:= \langle ||\vec{p}^*-\vec{p}||^2\rangle = \sum_{s=1}^{15}\langle \left(p_s^*-p_s\right)^2\rangle = \sum_{s=1}^{15}\Sigma^{(ss)}_{\vec{p}^*-\vec{p}} \\
& =\tr\Sigma_{\vec{p}^*-\vec{p}}=\tr(L^{-1}\Sigma L^{-T}),
\end{align}
where $\Sigma_{\vec{p}^*-\vec{p}}:=L^{-1}\Sigma L^{-T}$ is the covariance matrix of the random vector $\vec{p}^*-\vec{p}$ with matrix elements $\Sigma^{(su)}_{\vec{p}^*-\vec{p}}$. Note that $\langle\vec{p}^*-\vec{p}\rangle=\vec{0}$ since $\langle\vec{p}^*\rangle=\vec{p}$, which we have used in the second equality. 
We extend the definition of $\langle D^2\rangle$ to singular $L$ by defining it to be $\infty$ in this case.

Now that we have characterised the statistics of GSC and defined a quality measure, we are able to minimize the statistical error of GSC. In the next section, we will look for a set of sequences, initial states and measurements which achieves a small mean squared distance under the restrictions that, first, $L$ is a well-conditioned matrix and, second, the settings of GSC are physically implementable.

\section{Optimizing GSC}\label{s:optimizing}
In this section, we employ the statistical characterization we have developed in Sec.~\ref{s:statistics} to optimize the calibration protocol GSC. To do so, we introduce some parameters in the settings of GSC and optimize the statistical error over these parameters. Aiming for a physical implementation of GSC, we further modify the optimised settings in order to have a small number of different gates. Before all this, however, we start analysing the version of GSC proposed in Ref.~\cite{GSCpaper} with the quality measure defined in Sec.~\ref{s:statistics}.

In Ref.~\cite{GSCpaper}, GSC is introduced and applied in a two-qubit scenario to calibrate a perturbed $\cnot$ gate, as we have summarized in Sec.~\ref{s:GSC}. When we evaluate the statistical error of this version of the protocol in terms of the mean squared distance defined in Eq.~(\ref{e:D2}), we obtain that $\D\approx 7.4/N$.

As we have already mentioned in Sec.~\ref{s:statistics}, we can assume that each estimated response, $R_s^*$, is distributed according to a normal distribution such that $\mathcal{N}(R_s(\vec{p}),\Sigma)$ with $R_s(\vec{p})$ and $\Sigma$ defined in Eqns.~(\ref{e:Rs}) and (\ref{e:Sigma}), respectively. For the settings chosen in Ref.~\cite{GSCpaper}, each estimated response is distributed around zero when $\vec{p}=\vec{0}$ (see the last column of Tab.~\ref{t:sequences}). The calibration of the $\cnot$ is, therefore, performed at the point of maximal uncertainty, i.e., where the variance of each response reaches its maximum, $\Sigma_{ss}=1/N$, as we can see in Fig.~\ref{f:variance}. Here we have restricted the attention to the diagonal elements of $\Sigma$ because all settings are independent of each other.

We propose to slightly modify GSC in order to benefit from working at a point with smaller statistical error. To this end, we allow a different angle for each rotation of each GSC sequence so that the calibration of a $\tilde{\cnot}$ is performed using the sequences in Tab.~\ref{t:new_sequences25}. Our idea is to numerically optimize the mean squared distance of these settings over the angles $\theta_1,\dots,\theta_{25}$ with a local derivative-free optimizer that allows boundary constraints. 

\begin{figure}[htb]
\centering
\includegraphics[scale=0.6]{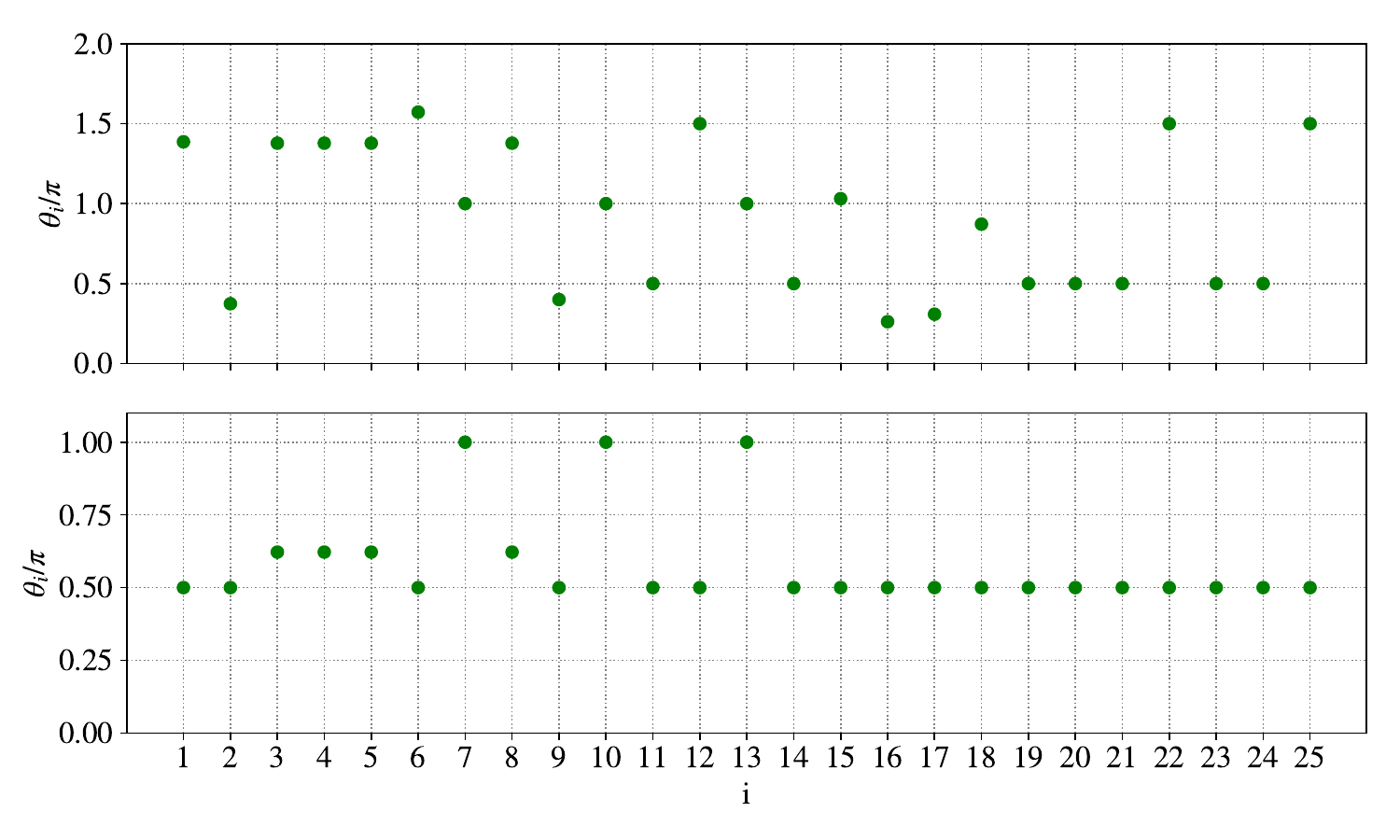}
\caption{Above, angles that minimize the mean squared distance of the calibration of a $\tilde{\cnot}$ using sequences in Tab.~\ref{t:new_sequences25} (Eq.~(\ref{e:D2})). Below, the same angles after a folding transformation (i.e., mapping $\theta_i\longrightarrow2\pi-\theta_i$ for all $\theta_i>\pi$) and setting $\theta_1=\theta_2=\theta_6=\theta_9=\theta_{15}=\theta_{16}=\theta_{17}=\theta_{18}=\pi/2$. Since the mean squared distance is invariant under folding transformation and independent of $\theta_1$, $\theta_2$, $\theta_6$, $\theta_9$, $\theta_{15}$, $\theta_{16}$, $\theta_{17}$, $\theta_{18}$, both angle combinations achieve the minimum value, $\D\approx3.4/N$.}
\label{f:angles25}
\end{figure}

We use the algorithm specified as \textsc{nlopt\_ln\_bobyqa} in the NLopt library \cite{NLOpt}, which, according to the documentation, is derived from the BOBYQA subroutine introduced in Ref.~\cite{optimizer}. This algorithm optimizes the objective function by iteratively constructing its quadratic approximation. 
We set boundary constraints such that $\theta_i\in[0,2\pi]$ for all $i=1,\dots,25$. A stopping tolerance for the optimizer is set to $10^{-10}$ at the angles.
Note that a local optimizer such as BOBYQA does not always converge to a global minimum. 
In order to gain confidence that our result is indeed a global minimum, we have repeated the optimization $10^4$ times with random starting points. 
In roughly one in $30$ runs, the optimizer stopped at values within a factor of $10^{-5}$ of the smallest mean squared error observed.

We have obtained a minimum mean squared distance of $\D\approx 3.4/N$ with the combination of angles in Fig.~\ref{f:angles25}. This means that one can achieve a reduction of the statistical error of GSC by a factor of $0.46$ with respect to the version of the GSC proposed in Ref.~\cite{GSCpaper}.

\begin{table}[phtb]
\begin{center}
{\renewcommand{\arraystretch}{2}
\begin{tabular}{|c|c|c|c|c|c|}
\hline
$s$ & 1st gate & 2nd gate & 3rd gate & 4th gate & $M_s$ \\ 
\hline 
1 & $\cnot$ & $\text{X}^{(1)}_{\theta_{15}}$ &  &  & $\tau_{12}$ \\ 
 
2 & $\text{X}^{(1)}_{\theta_1}$ & $\cnot$ &  &   & $\tau_{12}$ \\ 
 
3 & $\cnot$ & $\text{Y}^{(1)}_{\theta_{16}}$ &  &   & $\tau_{12}$ \\ 
 
4 & $\text{Y}^{(1)}_{\theta_2}$ & $\cnot$ &   &   & $\tau_{12}$ \\ 

5 & $\cnot$ & $\text{X}^{(2)}_{\theta_{17}}$ &   &   & $\tau_3$ \\ 

6 & $\cnot$ & $\text{Y}^{(2)}_{\theta_{18}}$ &   &   & $\tau_3$ \\ 

7 & $\text{X}^{(1)}_{\theta_3}$ & $\cnot$ & $\text{X}^{(1)}_{\theta_{19}}$ &  & $\tau_{12}$ \\ 

8 & $\text{X}^{(1)}_{\theta_4}$ & $\cnot$ & $\text{Y}^{(1)}_{\theta_{20}}$ &   & $\tau_{12}$ \\ 

9 & $\text{Y}^{(1)}_{\theta_5}$ & $\cnot$ & $\text{X}^{(1)}_{\theta_{21}}$ &   & $\tau_{12}$ \\ 

10 & $\text{X}^{(1)}_{\theta_6}$ & $\text{X}^{(2)}_{\theta_7}$ & $\cnot$ &   & $\tau_3$ \\ 

11 & $\text{Y}^{(1)}_{\theta_8}$ & $\cnot$ & $\text{Y}^{(1)}_{\theta_{22}}$ &   & $\tau_{12}$ \\ 

12 & $\text{Y}^{(1)}_{\theta_9}$ & $\text{Y}^{(2)}_{\theta_{10}}$ & $\cnot$ &   & $\tau_3$ \\ 

13 & $\text{Y}^{(2)}_{\theta_{11}}$ & $\cnot$ & $\text{X}^{(2)}_{\theta_{23}}$ &   & $\tau_3$ \\ 

14 & $\text{X}^{(1)}_{\theta_{12}}$ & $\cnot$ & $\cnot$ & $\text{Y}^{(1)}_{\theta_{24}}$ & $\tau_{12}$  \\

15 & $\text{X}^{(1)}_{\theta_{13}}$ & $\text{Y}^{(2)}_{\theta_{14}}$ & $\cnot$ & $\text{X}^{(2)}_{\theta_{25}}$ & $\tau_3$ \\ 
\hline
\end{tabular}}
\caption{List of settings to optimize the calibration of a $\tilde{\cnot}$ gate in Eq.~(\ref{e:E}) via GSC. A numerical optimization over $\theta_1,\dots,\theta_{25}$ allows to reduce the statistical error with respect to the GSC with sequences in Tab.~\ref{t:sequences}. For clarity, the linear responses of the measurements are in Appendix~\ref{a:responses}.}
\label{t:new_sequences25}
\end{center}
\end{table}

This new version of GSC requires sixteen single-qubit rotations for a maximal sequence depth of four gates. Specifically, we need one rotation for each different angle and, for some angles, we need two rotations: one rotation around the X-axis and one rotation around the Y-axis. Note that, although all rotations X$_{n\phi}$ and Y$_{n\phi}$, where $n$ is an integer, can be implemented by applying $n$ times the rotation X$_{\phi}$ or Y$_{\phi}$, respectively, this implies a cost in the sequence depth. 

From an experimental point of view, tuning and calibrating sixteen different single-qubit rotations is undesirable. Therefore, we would like to reduce the number of different angles without introducing a significant cost in the mean squared distance. We have observed numerically that the mean squared distance is invariant under the following transformations of the set of \emph{optimal} angles (the transformations are not symmetries of the objective function at non-optimal points):
\begin{enumerate}
\item
\emph{degeneracy}:
any change of the angles
$\theta_1$, $\theta_2$, $\theta_6$, $\theta_9$, $\theta_{15}$, $\theta_{16}$, $\theta_{17}$, and $\theta_{18}$
as long as $L$ remains non-singular;
\item \emph{global reflection}: $\theta_i \longrightarrow 2\pi-\theta_i$ for all $i=1,\dots,25$;
\item \emph{local reflection}: $\theta_i \longrightarrow 2\pi-\theta_i$ only for one $i\in\{1,\dots,25\}$ as long as $L$ remains non-singular;
\item \emph{folding transformation}: $\theta_i\longrightarrow2\pi-\theta_i$ for all $\theta_i>\pi$.
\end{enumerate}
It would be tempting to set all the angles listed under 1.\ to zero, thereby eliminating the corresponding gates.
However, it turns out that this choice does lead to a singular $L$ and will therefore not be considered here.
Still, a significant reduction of complexity can be reduced by making use of the above symmetries.
The folding transformation allows us to reduce, without any cost in the mean squared distance, the number of different rotations to thirteen for a maximal sequence depth of four gates.
At the price of increasing the sequence depth to seven gates, one can further reduce to eleven distinct rotations.

\begin{figure}[htb]
\centering
\includegraphics[scale=0.4]{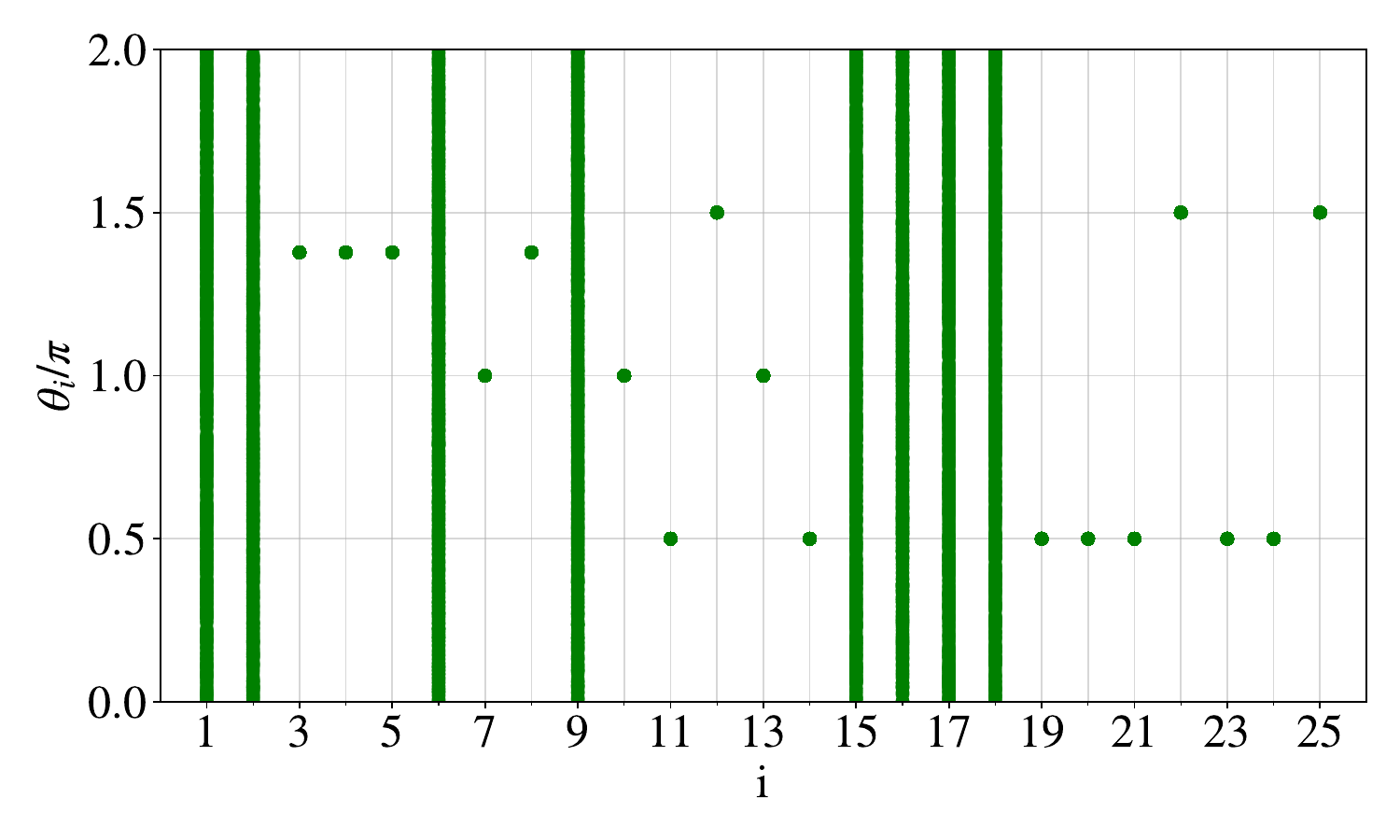}
\caption{Combinations of angles that give a mean squared distance $\D\approx3.4/N$. The mean squared distance is independent of $\theta_1$, $\theta_2$, $\theta_6$, $\theta_9$, $\theta_{15}$, $\theta_{16}$, $\theta_{17}$, $\theta_{18}$, and thus we can set these angles at any value in $[0,2\pi]$ without any cost in the statistical error.}
\label{f:degeneracy}
\end{figure}

To reduce even more the number of gates in the set of available gates, we use the fact that the minimal mean squared distance is degenerate. In Fig.~\ref{f:degeneracy}, we can see that the minimum mean squared distance is achieved regardless of the value of the angles $\theta_1$, $\theta_2$, $\theta_6$, $\theta_9$, $\theta_{15}$, $\theta_{16}$, $\theta_{17}$ and $\theta_{18}$. As mentioned before, we can unfortunately not set these angles simply to zero, and thus get rid off the corresponding rotations, because $L$ becomes singular. These angles can alternatively be set to $\pi/2$, which is a single-qubit gate already required to implement GSC (see Fig.~\ref{f:angles25}). The mean squared distance remains at $\D\approx3.4/N$ and now GSC needs only four rotations and a maximal sequence depth of five gates.

Ultimately, we are proposing a GSC protocol that calibrates a two-qubit $\tilde{\cnot}$ gate with a statistical error $0.46$ times smaller than in Ref.~\cite{GSCpaper}. The set of allowed gates is 
\begin{align*}
    \{\cnot,\text{X}^{(1)}_{\pi/2},\text{Y}^{(1)}_{\pi/2},\text{X}^{(2)}_{\pi/2},\text{Y}^{(2)}_{\pi/2},\text{X}^{(1)}_\theta,\text{Y}^{(1)}_\theta\},
\end{align*} 
where $\theta=0.62208\pi$, and the maximal sequence depth is five, as we can see in Tab.~\ref{t:new_sequences}.

\begin{table}[phtb]
\begin{center}
{\renewcommand{\arraystretch}{2}
\begin{tabular}{|c|c|c|c|c|c|c|c|}
\hline
$s$ & 1st gate & 2nd gate & 3rd gate & 4th gate & 5th gate & $M_s$ & $R_s(\vec{p})$ \\ 
\hline 
1 & $\cnot$ & $\text{X}^{(1)}_{\pi/2}$ &  &  &  & $\tau_{12}$ & $0 - 2 p_5 + 2 p_{10}$ \\ 
 
2 & $\text{X}^{(1)}_{\pi/2}$ & $\cnot$ &  &  &   & $\tau_{12}$ & $0 - 2 p_4 - 2 p_7$ \\ 
 
3 & $\cnot$ & $\text{Y}^{(1)}_{\pi/2}$ &  &  &   & $\tau_{12}$ & $0 - 2 p_6 - 2 p_9$ \\ 
 
4 & $\text{Y}^{(1)}_{\pi/2}$ & $\cnot$ &   &  &   & $\tau_{12}$ & $0 - 2 p_8 -2 p_{11}$ \\ 

5 & $\cnot$ & $\text{X}^{(2)}_{\pi/2}$ &   &  &   & $\tau_3$ & $0 - 2 p_1 - 2 p_{13}$ \\ 

6 & $\cnot$ & $\text{Y}^{(2)}_{\pi/2}$ &   &  &   & $\tau_3$ & $0 - 2 p_2 -2 p_{14}$ \\ 

7 & $\text{X}^{(1)}_{\theta}$ & $\cnot$ & $\text{X}^{(1)}_{\pi/2}$ &  &   & $\tau_{12}$ & $0 - 2 \sin \theta p_2 - 2 \cos \theta p_5  + 2 p_{10}$ \\ 

8 & $\text{X}^{(1)}_{\theta}$ & $\cnot$ & $\text{Y}^{(1)}_{\pi/2}$ &  &   & $\tau_{12}$ & $0 - 2 p_6 - 2 \cos \theta p_9 - 2 \sin \theta p_{13}$ \\ 

9 & $\text{Y}^{(1)}_{\theta}$ & $\cnot$ & $\text{X}^{(1)}_{\pi/2}$ &  &   & $\tau_{12}$ & $0 - 2 \cos \theta p_5 + 2 p_{10} + 2 \sin \theta p_{13}$ \\ 

10 & $\text{X}^{(1)}_{\pi/2}$ & $\text{X}^{(2)}_{\pi/2}$ & $\text{X}^{(2)}_{\pi/2}$ & $\cnot$ &   & $\tau_3$ & $0 + 2 p_4 - 2 p_7$ \\ 

11 & $\text{Y}^{(1)}_{\theta}$ & $\cnot$ & $\text{Y}^{(1)}_{\pi/2}$ &  &  & $\tau_{12}$ & $0 - 2 \sin \theta p_2 - 2 p_6 - 2 \cos \theta p_9$ \\ 

12 & $\text{Y}^{(1)}_{\pi/2}$ & $\text{Y}^{(2)}_{\pi/2}$ & $\text{Y}^{(2)}_{\pi/2}$ & $\cnot$ &   & $\tau_3$ & $0 + 2 p_8 - 2 p_{11}$ \\ 

13 & $\text{Y}^{(2)}_{\pi/2}$ & $\cnot$ & $\text{X}^{(2)}_{\pi/2}$ &  &  & $\tau_3$ & $0 + 2 p_3 + 2 p_{15}$ \\ 

14 & $\text{X}^{(1)}_{\pi/2}$ & $\cnot$ & $\cnot$ & $\text{Y}^{(1)}_{\pi/2}$ &  & $\tau_{12}$ & $0 - 2 p_3 -4 p_{12} - 2 p_{15}$ \\ 

15 & $\text{X}^{(1)}_{\pi/2}$ & $\text{X}^{(1)}_{\pi/2}$ & $\text{Y}^{(2)}_{\pi/2}$ & $\cnot$ & $\text{X}^{(2)}_{\pi/2}$ & $\tau_3$ & $0 - 2 p_3 + 2 p_{15}$ \\ 
\hline
\end{tabular}}
\caption{List of settings to reduce the statistical error in the calibration of a $\tilde{\cnot}$ using GSC. The set of allowed gates is $\{\cnot,\text{X}^{(1)}_{\pi/2},\text{Y}^{(1)}_{\pi/2},\text{X}^{(2)}_{\pi/2},\text{Y}^{(2)}_{\pi/2},\text{X}^{(1)}_\theta,\text{Y}^{(1)}_\theta\}$ with $\theta=0.62208\pi$.}
\label{t:new_sequences}
\end{center}
\end{table}

\subsection{Introducing imperfect measurements} \label{s:imperfect}

In all previous sections, the measurements performed in GSC are assumed to be perfect. 
In other words, we have assumed that one can perform a projective measurement in some orthogonal basis.
However, in the laboratory, devices are imperfect and subject to noise.
Here, we investigate how the results behave under a simple noise model.

Let us consider the scenario presented in Sec.~\ref{s:GSC}: calibrating a perturbed gate, $\tilde{\cnot}$, (see Eq.~(\ref{e:E})) using GSC. All measurements performed in GSC are Pauli observables, and thus they have only two outcomes, the positive and the negative outcome.
The error model used here is given by an perfect Pauli measurement, followed by a \emph{asymmetric binary channel}.
More precisely, the model takes two error parameters $F_+, F_1 \in [0,1]$ such that if the perfect measurement yields the outcome $+1$, the imperfect one will return $+1$ with probability $F_+$ and $-1$ with probability $1-F_+$.
The situation for the $(-1)$-outcome of the perfect Pauli measurement is analogous.

The probability to read a positive outcome on the measurement device after measuring $M_s$ is then
\begin{align*}
Q_s(+):= & F_+q_s+(1-F_-)(1-q_s),\\
= & \dfrac{1}{2}\left[1+F_+-F_-+R_s\left(F_++F_--1\right)\right],
\end{align*}
where $q_s$ is the probability of obtaining the positive outcome, i.e., $q_s=(1+R_s)/2$, and $R_s$ is defined in Eq.~(\ref{e:Rs}). Recall that $R_s$ depends on the error parameters, and so do $q_s$ and $Q_s(+)$.

The probability that the measurement device gives a negative outcome after measuring $M_s$ is
\begin{align*}
Q_s(-)= & F_-(1-q_s)+(1-F_+)q_s \\
= & \dfrac{1}{2}\left[1+F_--F_+-R_s\left(F_++F_--1\right)\right].
\end{align*}
It is easy to check that $Q_s(+)+Q_s(-)=1$.

The \emph{imperfect measurement responses} are hence given by
\begin{align*}
\tilde{R}_s(\vec{p}) & = Q(+) - Q(-), \\
& = F_+-F_-+R_s(F_++F_--1).
\end{align*}
Then, the matrix elements of $L$ can be written as
\begin{align*}
\tilde{L}_{ru} & =\left.\dfrac{\partial \tilde{R}_r(\vec{p})}{\partial p_u}\right|_{\vec{p}=\vec{0}}, \\
& = (F_++F_--1)L_{ru}.
\end{align*}

As we have done in Sec.~(\ref{s:statistics}), we use the mean squared distance as measure of uncertainty to compare different versions of GSC. The mean squared distance considering imperfect measurements becomes
\begin{equation}\label{e:D2tilde}
\tilde{\D} =\tr(\tilde{L}^{-1}\tilde{\Sigma} \tilde{L}^{-T}),
\end{equation}
where the matrix elements of the covariance matrix of the perturbed measurement responses, $\tilde{\Sigma}$, are $\tilde{\Sigma}_{su}=\dfrac{1}{N}(1-\tilde{R}_s^2)\delta_{su}.$

Let $F_+=0.99$ and $F_-=0.98$ and consider the GSC presented in Ref.~\cite{GSCpaper}, which uses the sequences in Tab.~\ref{t:sequences}. The statistical error measured by the mean squared distance in Eq.~(\ref{e:D2tilde}) is $\tilde{\D}\approx7.8/N$.

\begin{figure}[htb]
\centering
\includegraphics[scale=0.55]{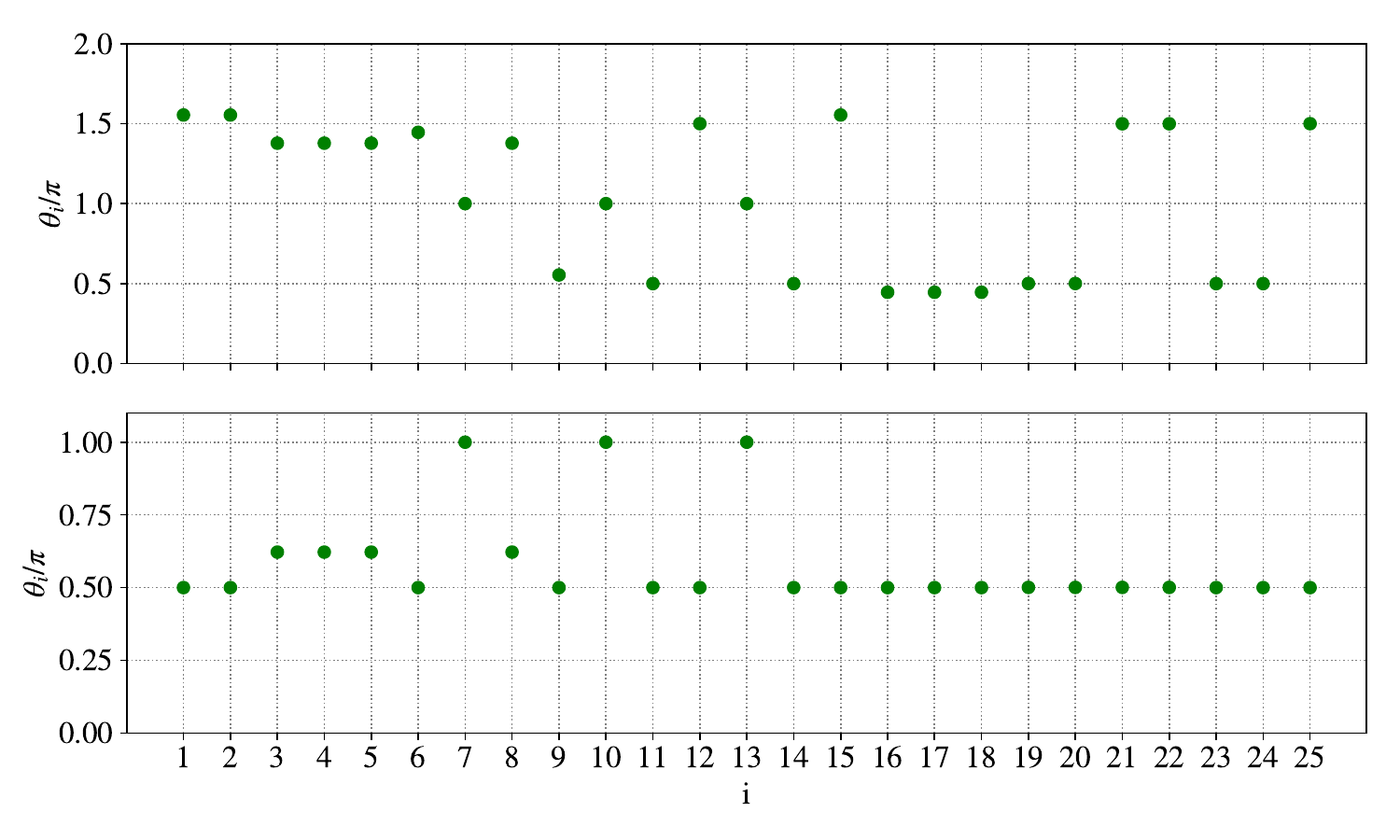}
\caption{Above, angles that minimize the mean squared distance in Eq.~(\ref{e:D2tilde}), which considers imperfect measurements in the calibration of a $\tilde{\cnot}$ using sequences in Tab.~\ref{t:new_sequences25}. Below, the same angles after a folding transformation (i.e., mapping $\theta_i\longrightarrow2\pi-\theta_i$ for all $\theta_i>\pi$) and setting $\theta_1=\theta_2=\theta_6=\theta_9=\theta_{15}=\theta_{16}=\theta_{17}=\theta_{18}=\pi/2$.}
\label{f:angles25F}
\end{figure}

Allow now a different angle for each GSC rotation, i.e., consider the sequences in Tab.~\ref{t:new_sequences25}. With the optimization procedure described in Sec.~\ref{s:optimizing}, we minimize the mean squared distance in Eq.~(\ref{e:D2tilde}) over the angles $\theta_1,\dots,\theta_{25}$. The angles in Fig.~\ref{f:angles25F} achieve a minimal mean squared distance at $\tilde{\D}\approx3.6/N$. This means a statistical error $0.46$ times smaller than in Ref.~\cite{GSCpaper}, but requires seventeen different rotations with a maximal sequence depth of four gates. As in the case of perfect measurement, the number of rotations can be further reduced to thirteen at the cost of a sequence depth of seven gates.

In analogy with the case with perfect measurements treated in Sec.~\ref{s:optimizing}, the number of rotations can be reduced by considering symmetries of the optimal mean squared distance. We have numerically shown that the minimal $\tilde{\D}$ is invariant under folding transformation as well as global and local reflection of the \emph{optimal} angles. After considering the folding transformation, the number of rotations reduces to twelve without any change in the mean squared distance and with a maximal sequence depth of four gates.

If twelve different single-qubit rotations are undesirable to perform in the lab, we can set the angles $\theta_1=\theta_2=\theta_6=\theta_9=\theta_{15}=\theta_{16}=\theta_{17}=\theta_{18}=\pi/2$. In contrast to the case of perfect measurements, the mean squared distance for imperfect measurements is not degenerate. Setting these angles implies, however, an insignificant increase in the statistical error. In other words, we obtain a $\tilde{\D}\approx3.6/N$, which means a reduction of the statistical error by a factor of $0.46$ with respect to the statistical error in Ref.~\cite{GSCpaper},  as mentioned above. Note that the final settings proposed for GSC are the same in both cases, considering perfect and imperfect measurements.

In summary, we propose to calibrate a perturbed $\cnot$ gate using GSC with the sequences in Tab.~\ref{t:new_sequences} and $\theta=0.62208\pi$. This version of GSC requires a maximal sequence depth of five gates and a set of allowed gates consisting of $\{\cnot,\text{X}^{(1)}_{\pi/2},\text{Y}^{(1)}_{\pi/2},\text{X}^{(2)}_{\pi/2},\text{Y}^{(2)}_{\pi/2},\text{X}^{(1)}_\theta,\text{Y}^{(1)}_\theta\}$, which contains two more single-qubit gates than the set proposed in Ref.~\cite{GSCpaper}. Nevertheless, it reduces the statistical error by a factor of $0.46$ for perfect measurements as well as for imperfect measurements with $F_+=0.99$ and $F_-=0.98$. All results are summarized in Tab.~\ref{t:summary}.

\begin{table}[htb]
\begin{center}
\begin{tabular}{|c|c|c|c|c|}
\hline 
 & $[F_+,F_-]$ & $\D\cdot N$ & $\#$ rotations & Maximal sequence depth\\
\hline 
Angles of Ref.~\cite{GSCpaper} & $[1,1]$ & $7.4$ & $2$  & 4 \\
\hline 
Optimal angles & $[1,1]$ & $3.4$ & $17$ & 4 \\ 
\hline 
Folded angles & $[1,1]$ & $3.4$ & $12$ & 4 \\ 
\hline 
Modified angles & $[1,1]$ & $3.4$ & $4$ & 5 \\
\hline\hline 
Angles of Ref.~\cite{GSCpaper} & $[0.99,0.98]$ & $7.8$ & $2$ & 4 \\
\hline
Optimal angles & $[0.99,0.98]$ & $3.6$ & $17$ & 4 \\ 
\hline 
Folded angles & $[0.99,0.98]$ & $3.6$ & $12$ & 4 \\ 
\hline 
Modified angles & $[0.99,0.98]$ & $3.6$ & $4$ & 5\\
\hline 
\end{tabular}
\caption{Summary of numerical results, where the statistical error is measured by the mean squared distance, $\D$, and the forth column shows the number of single-qubit rotations required.}
\label{t:summary}
\end{center}
\end{table}

\section{Simulation} \label{s:simulation}
The original GSC protocol was designed with the in-situ tuneup of GaAs based $S-T_0$ qubits in mind. In this section, we perform simulations of the refined GSC algorithm on this qubit type and validate its convergence behaviour following Ref.~\cite{GSCpaper}.

We follow the overall model presented in Ref.~\cite{Cerfontaine2020h}. Spin qubits in GaAs are created by placing metal electrodes on top of a semiconductor heterostructure, allowing for local control of the chemical potential and the exchange interactions ($J_{j,j+1}, j \in \{ 1,2,3 \}$) within an array of four linear adjacent quantum dots. Each site is occupied by one electron, whose spin experiences the magnetic field $B_j, j \in \{ 1,2,3,4 \}$ which can be different for each site. This results in the Hamiltonian, with $\hbar = 1$,
\begin{align}
\label{eq:hj}
	H = & \sum_{j=1}^3 \frac{J_{j,j+1}}{4}\vecsig ^{(j)} \cdot \vecsig ^{(j+1)} + \frac{1}{2}  \sum_{j=1}^4 B_j \sigma^{(j)}_z,
\end{align}
where $\sigma^{(j)}$ is the Pauli operator acting on the spin on site $j$. The $\{ \ket{\mathbin{\uparrow\downarrow}},\ket{\mathbin{\downarrow\uparrow}}\}\otimes \{ \ket{\mathbin{\uparrow \downarrow}},\ket{\mathbin{\downarrow \uparrow}}\}$ subspace is chosen as the qubit encoding. 

The states $\ket{\mathbin{\uparrow \uparrow \downarrow \downarrow}}$ and $\ket{\mathbin{\downarrow \downarrow \uparrow \uparrow}}$ are considered leakage states but their occupation can be largely suppressed by choosing $J_{23}/|b_{23}| \ll 1$, where $b_{j,j+1} = B_j - B_{j+1}$ is the difference in local magnetic field between neighboring sites. The gradients achievable in experiments using dynamic nuclear polarization are within the range $0.1\:$ns$^{-1}$ to $7\:$ns$^{-1}$ and are chosen to be $b_{23}=7\:$ns$^{-1}$ and  $b_{12} = -b_{34} =1\:$ns$^{-1}$ in this simulation. The exchange interaction $J_{j,j+1}$ is related to the detuning voltages between neighboring sites using the phenomenological relation $J_{j,j+1}(\epsilon_{j,j+1})=J_0\exp{(\epsilon_{j,j+1}/\epsilon_0)}$, with $\epsilon_0=0.272$ mV and $J_0=1$ ns$^{-1}$
\cite{Dial}. The detuning voltages can be controlled using arbitrary waveform generators with a sample rate of about 1 GS/s, with the exact value depending on the specific hardware used. Each gate applied to the qubit is therefore parameterized in three detuning sequence with typical length of $N_\mathrm{samp}=20$ to $50$ samples.

These sequences are convoluted with a typical impulse response of the experimental setup \cite{Cerfontaine2020h}. Hence we can approximate the time-dependence of each $J_{j,j+1}(t)$ and numerically calculate the time evolution of the two qubit system using a piece-wise constant approximation. Using the unitary obtained in this fashion we can quantify the infidelity by comparing to the unitary of the desired operation $U_t$, which is then given by $\mathcal{I}_\mathrm{u}(\epsilon)=1-\mathcal{F}_\mathrm{u}[  U(\epsilon), U_t ]$, where $\mathcal{F}_\mathrm{u}$ is the fidelity of the operation \cite{Nielsen2002a}. Similarly the leakage out of the computational subspace is given by $\mathcal{L} = 1 - \tr{(V_c^\dagger V_c/4)}$, where $V_c$ is the truncation of $U(\epsilon)$ into the computational subspace. 

We also take into account effects of charge noise affecting $\epsilon$ and hyperfine noise affecting $b_{j,j+1}$ on the fidelity and leakage by averaging them over 100 computer generated noise realizations. Hyperfine noise in GaAs is assumed to have a standard deviation of $\sigma_b = 0.3$\:mT when stabilized using dynamic nuclear polarization \cite{Cerfontaine2020}. For charge noise we use a combination of quasistatic noise with $\sigma_\epsilon = 8$\:\textmu V and white noise with a strength of $4\times 10^{-20}$\:V$^2$\,Hz$^{-1}$ following experimentally determined values \cite{Dial}. 

We use the initial state $\hat{\rho}_i =\ket{\mathbin{\uparrow \downarrow \uparrow \downarrow}}\bra{\mathbin{\uparrow \downarrow\uparrow \downarrow}} = \ket{00} \bra{00}$ native to this qubit type \cite{Cerfontaine2020h}. The inability to discern the three possible triplet states leads to the following measurement operators $\hat{M}_1$ and $\hat{M}_2$ \cite{Barthela}, \cite{Barthelb} when the basis states $(\ket{\downarrow \downarrow \uparrow \uparrow},
\ket{\uparrow \downarrow \uparrow \downarrow},
\ket{\uparrow \downarrow \downarrow \uparrow},
\ket{\downarrow \uparrow \uparrow \downarrow},
\ket{\downarrow \uparrow \downarrow \uparrow},
\ket{\uparrow \uparrow \downarrow \downarrow })$ are taken into account:
\begin{equation*}
\hat{M}_1 =
    \begin{pmatrix}
-1 &   &   \\
  &  \sigma_3 \otimes \sigma_0 &   \\
  &   &  -1 \\
\end{pmatrix},\,
\hat{M}_2 =
    \begin{pmatrix}
-1 &  & \\
&  \sigma_0 \otimes \sigma_3 & \\
&  &  -1 \\
\end{pmatrix}.
\end{equation*}

\begin{figure}[htb]
\centering
\includegraphics[scale=1]{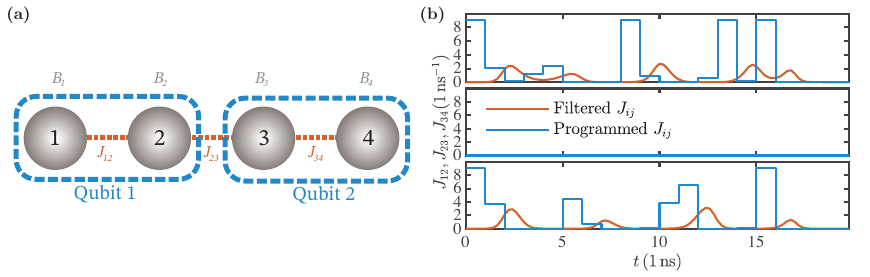}
\caption{\textbf{(a)} Sketch of the qubit system as implemented for the simulation. \textbf{(b)} Pulse sequence for a $X_{\theta}$ gate to be programmed on the AWG (blue) and seen by the qubit (orange). Exchange between the two qubit is turned off for this simulation ($ J_{23}=0 $) to reduce complexity. Figure adapted from Ref.~\cite{Cerfontaine2020h}.} 
\label{f:gate}
\end{figure}

For  the new GSC protocol we need the full gate set $\{\cnot,\text{X}^{(1)}_{\pi/2},\text{Y}^{(1)}_{\pi/2},\text{X}^{(2)}_{\pi/2},\text{Y}^{(2)}_{\pi/2},\text{X}^{(1)}_\theta,\text{Y}^{(1)}_\theta\}$ given above. Physical realizations assuming the above model have already been proposed in Ref.~\cite{Cerfontaine2020h} for the first five gates in this gate set and used to validate the original GSC sequences. We use the same numerical optimization from Ref.~\cite{Cerfontaine2020h} to obtain realizations of the additional $\text{X}^{(1)}_\theta$ and $\text{Y}^{(1)}_\theta$ gates. The pulse sequence for the former is shown in Fig.~\ref{f:gate}, where the programmed sequence is shown in addition to the exchange as seen by the qubit after filtering effects of the cryostat wiring and AWG bandwidth. A summary of the performance of all gates used in the following simulations are given in Tab.~\ref{t:best_xtheta_ec0}. 

Since the sequences in Tab.~\ref{t:new_sequences} are designed to calibrate the CNOT gate only, we leave all other gates in the gate set fixed. We perturb the $\epsilon$ sequences of the perfect CNOT by random voltages in order to obtain a set of starting values with initial infidelities $\mathcal{I}_\mathrm{initial}$ of up to 20\:\% due to systematic errors. We can now numerically evaluate and minimize $|\Delta R|=| \vec{R}(\vec{\epsilon})-\vec{R}_0 | = | \vec{R}(\vec{p})-\vec{R}_0 |$, where $R_0$ is the outcome of the sequences with perfect gates, assuming that our error parameters $\vec{p}$ are related to our experimental parameters $\vec{\epsilon}$ with some unknown transformation. We bound the detuning to values achievable in experiments $-5.4\epsilon_0<\epsilon<2.4\epsilon_0$ and perform the optimization numerically using the Levenberg-Marquardt algorithm. In each iteration we use the suggested values of $\epsilon$ to construct a unitary using the procedure lined out before and calculate the measurement responses when applying the GSC sequences. 

\begin{figure}[htb]
\centering
\includegraphics[scale=1]{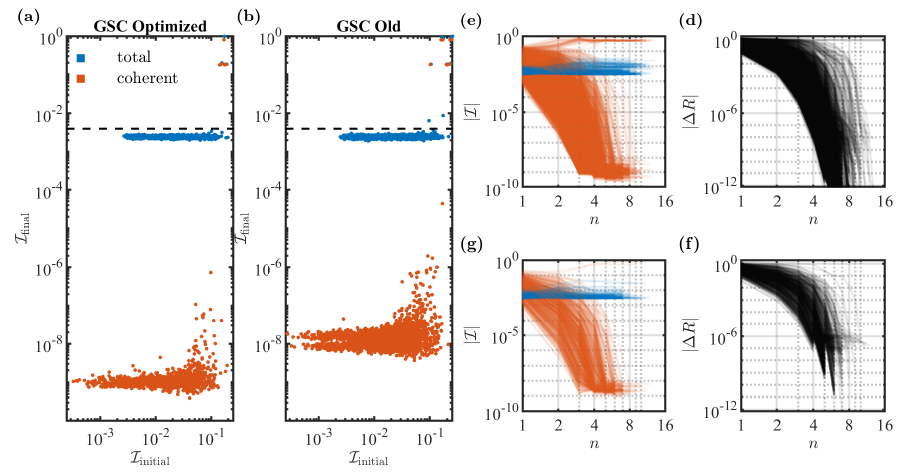}
\caption{Comparison of original and optimized GSC convergence in simulations. (a) and (b) show the infidelity after the optimization procedure for both sets of sequences. Infidelity (e) and sum of error syndromes (d) for the improved GSC sequences show a similar convergence behaviour as the original sequences shown in panels (g) and (f).  }
\label{f:result}
\end{figure}

We then optimize these responses as a function of the detunings $\vec{\epsilon}$. In order to control decoherence, in each iteration the infidelity of the CNOT gate is calculated and mapped to a depolarizing channel \cite{Cerfontaine2020h}. Two additional sequences $G_\mathrm{dec} = \text{CNOT} \times \text{CNOT} = \mathds{1}$ followed by a measurement on one of the qubits are performed at each iteration. 
Optimizing towards the measurement outcome without decoherence $R_s(\vec{p})=1$ allows us to incentivize the optimization towards results with minimal decoherence. 
Fig.~\ref{f:result} shows the final infidelity $\mathcal{I}_\mathrm{final}$ following the GSC procedure as a function of the initial infidelity $\mathcal{I}_\mathrm{initial}$ after the errors are introduced for the optimized (a) and for the original (b) GSC procedure. The optimized sequences perform as well or better then the original ones. The level reached for the final infidelity is slightly better for the optimized procedure as the new gates  $\text{X}^{(1)}_\theta$ and $\text{Y}^{(1)}_\theta$ have slightly better infidelities then the $\text{X}^{(1)}_{\pi/2}$ and $\text{Y}^{(1)}_{\pi/2}$ used for the original sequences. Systematic (unitary) errors can reliably be eliminated and the final infidelity is limited by incoherent errors due to noise. Leakage is still strongly suppressed after the GSC procedure due to the choice of $b_{23}$. The optimized sequences need the same number of iterations to converge as the original ones as can be seen from the error syndromes and the fidelity as a function of iteration number in Fig.~\ref{f:result} (e)-(f). Since the convergence behaviour remains unchanged with the optimized sequences, we conclude that a careful experimental design can lead to a significant reduction in measurement time caused by the decrease in averaging needed under realistic conditions.

\section{Experimental Considerations} \label{s:experiment}

In the following we give a brief introduction in the implementation of GSC on the physical experiment. Pulses are executed on an arbitrary waveform generator connected to dedicated high frequency (RF) gates on the qubit chip using coaxial lines. After each pulse the qubit is parked at its predetermined measurement position and a measurement is triggered for a certain amount of time (usually about 4\:\textmu s). Pulses are grouped into scanlines which are executed together on the AWG. Since the nuclear magnetic field needs to be constantly stabilized in GaAs based systems, GSC pulses are interleaved with DNP pulses keeping the Overhauser field gradients between the spins at the desired level. Typical scanlines will contain about 1000 pulses with about half of them for DNP and GSC respectively. Scanlines are usually repeated about 10 times directly on the AWG in order to reduce the overhead.
The time in between these groups of scanlines is used for additional tasks such as data transfer, processing and plotting. The process of executing a scanline $10\times$ and performing all the necessary post-processing took about 1\:s on past experiments with a pure measurement time of 40\:ms. In order to gain the necessary accuracy for GSC this entire process is then repeated about 100 times, with earlier iterations requiring a less  accuracy and therefore less averaging. Therefore, a typical GSC iteration takes a few tenths of seconds of wall time with only about 4\:\% of actual measurement time. Although it could be improved in principle, this discrepancy has an advantage in GaAs. 
While being approximately quasistatic on the scale of a single gate duration, the stabilized nuclear spin field is still subject to variation on the scale of the repetition time of DNP cycles. This introduces a systematic deviation from the programmed field gradient with a distribution of about $0.3\:$mT in width for each gate. 
As the starting points for the GSC algorithm are numerically optimized to decouple from the nuclear field for a given mean of the distribution, a longer runtime of the algorithm translates to a more exhaustive sampling of the field distribution, which in turn ensures that the dynamical decoupling properties of the starting gates is retained throughout the calibration procedure.

\begin{figure}[htb]
\centering
\includegraphics[scale=.8]{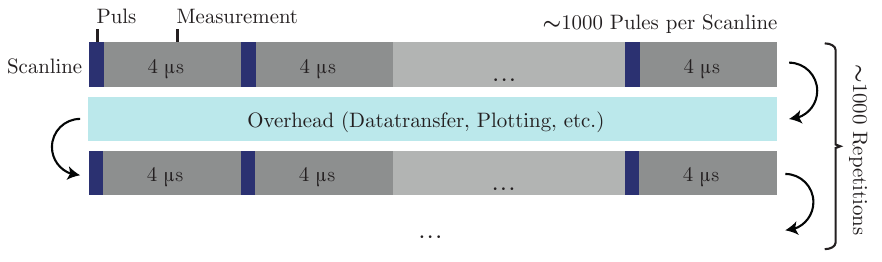}
\caption{Schematic of how GSC experiments are performed. Pulses are grouped into scanlines that are executed on an AWG with simultaneous data acquisition. Each scanline contains a mix of DNP and GSC pulses. Pulsetimes are usually around $10\:$ns to $100\:$ns. Time between scanlines is used for tasks such as data transfer and plotting. A usual experiment contains about 1000 scanlines depending on the required accuracy.}
\label{f:measurement}
\end{figure}

While this seems to contradict the proposed speedup of the GSC procedure using the optimized sequences in Tab.~\ref{t:new_sequences}, we would like to point out that these sequences will still reduce the required averaging time (the outermost repetitions in Fig.~\ref{f:measurement} significantly). This means that in the same amount of wall time we can tune an even larger number of gates simultaneously, outweighing the increased number of single qubit gates in our new gate set. Additionally, we would like to point out that this issue is only relevant for spin qubits in GaAs, where DNP is necessary. GSC itself, and our optimized protocol in particular, can be deployed independent of the specific qubit type. Assuming that computational overhead in between measurements can be reduced by performing tasks in parallel or using specialized hardware, time saved due to lower averaging caused by optimized statistical error can translate directly to faster GSC calibration.

Considering purely measurement time the above algorithm will lead to a 2.1-fold improvement in needed measurement time for the CNOT optimization, at the cost of two additional single qubit gates that need to be optimized beforehand. Assuming these need to be tuned for each qubit individually, we need a set of 6 sequences for each pair. Without an improvement due to design of experiment for the single qubit gates, this leads to a pure measurement time of $40\:$ms $\times$ 6 sequences $\times$ 10 iterations $= 2.4\:$s for each gate pair, using a typical iteration count observed in experiments. For the CNOT we get $40\:$ms $\times$ 15 sequences $\times$  10 iterations $= 6.0\:$s and $2.76\:$s with the optimized sequences. This allows us to compare the total time for the improved GSC with 6 single-qubit gates and the CNOT giving $3 \times 2.4\:$s $+ 2.76\:$s $= 9.96\:$s to the time needed without improvements $2 \times 2.4\:$s $+ 6.0\:$s $= 10.8\:$s showing a clear improvement in run time. This effect should be even more prominent if single qubit gates are also subject to improved sequences or tuned self consistently in a set of sequences that is designed with the above considerations in mind.

\section{Conclusions and outlook} \label{s:outlook}
We have seen that optimising the design of an experiment is crucial for minimising statistical errors. In particular, we have proposed a new version of GSC that reduces the statistical error by a factor of $0.46$ for perfect measurements as well as for imperfect measurements. To achieve this, one initial state, two Pauli measurements and six single-qubit rotations must be available.

Several questions remain open. Although we have managed to reduce the statistical error here, a physical explanation of the optimal sequences is still missing. We also do not know why the mean squared distance is degenerate for perfect measurements and why the degeneracy breaks down for imperfect measurements (even when $F_+=F_-\neq1$).

Our optimisation has been based on parameterising the GSC sequences, but different initial states and measurements could also be considered. Furthermore, we have used fifteen settings to solve fifteen error parameters. Considering additional settings, however, and making a statistical treatment of the redundant information could further reduce the statistical error.

Finally, it would also be interesting to put our results into practice, and thus experimentally calibrate a $\tilde{\cnot}$ with the version of GSC proposed in this article.

\section{Acknowledgments}
Funded by the Deutsche Forschungsgemeinschaft (DFG, German Research Foundation) under Germany's Excellence Strategy - Cluster of Excellence Matter and Light for Quantum Computing (ML4Q) EXC 2004/1-390534769.

\section{Data availability}
The data utilized in this research paper is openly accessible and can be freely obtained in \cite{dataset}. We encourage fellow researchers to utilize this dataset to replicate, validate, and expand upon the findings presented in our study.

\newpage

\appendix

\section{Supplementary material}

\subsection{Responses} \label{a:responses}
In Sec.~\ref{s:optimizing}, we have considered GSC with the settings summarized in Tab.~\ref{t:new_sequences25}. We present here the measurement responses corresponding to this protocol.

Using Eqns.~(\ref{e:Rs}) and (\ref{e:Rsapprox}), we obtain that the responses up to first order in the error parameters are

\begin{equation*}
\begin{aligned}
R_1(\vec{p}) & = \cos\theta_{15} - 2\cos\theta_{15} p_5 + 2 \cos\theta_{15} p_{10}, \\
R_2(\vec{p}) & = \cos\theta_1 - 2 \cos\theta_1 p_4 - 2 \cos\theta_1 p_7, \\
R_3 (\vec{p}) & = \cos\theta_{16} - 2 \cos\theta_{16} p_6 - 2 \cos\theta_{16} p_9, \\
R_4(\vec{p}) & = \cos\theta_2 - 2 \cos\theta_2 p_8 - 2 \cos\theta_2 p_{11}, \\
R_5(\vec{p}) & = \cos\theta_{17} - 2 \cos\theta_{17} p_1 - 2 \cos\theta_{17} p_{13}, \\
R_6(\vec{p}) & = \cos\theta_{18} - 2 \cos\theta_{18} p_2 - 2 \cos\theta_{18} p_{14},\\
R_7(\vec{p}) & = \cos\theta_{19} \cos\theta_3 - 2 \cos\theta_{19} \cos\theta_3 p_2 - 2 \cos\theta_{19} \cos\theta_3 p_4 \\
& \quad \quad - 2 \cos\theta_3 \cos\theta_{19} p_5 - 2 \cos\theta_{19} \cos\theta_3 p_7 + 2 \cos\theta_{19} p_{10},\\
R_8(\vec{p}) & = \cos\theta_{20} \cos\theta_{4} - 2 \cos\theta_{20} \cos\theta_4 p_4 - 2 \cos\theta_{20} p_6 \\
& \quad \quad - 2 \cos\theta_{20} \cos\theta_4 p_7 - 2 \cos\theta_4 \cos\theta_{20} p_9 - 2 \cos\theta_{20} \cos\theta_4 p_{13}, \\
R_9(\vec{p}) & = \cos\theta_{21} \cos\theta_5 - 2 \cos\theta_5 \cos\theta_{21} p_5 - 2 \cos\theta_{21} \cos\theta_5 p_8 \\
& \quad \quad  + 2 \cos\theta_{21} p_{10} - 2 \cos\theta_{21} \cos\theta_5 p_{11} + 2 \cos\theta_{21} \cos\theta_5 p_{13}, \\
R_{10}(\vec{p}) & = \cos\theta_6 \cos\theta_7 - 2 \cos\theta_6 p_7 - 2 \cos\theta_7 \cos\theta_6 p_4 \\
& \quad \quad  - 2 \sin\theta_{7} p_{13} - 2 p_1 \cos\theta_6 \sin\theta_7, \\
R_{11}(\vec{p}) & = \cos\theta_{22} \cos\theta_8 - 2 \cos\theta_{22} p_6 - 2 \cos\theta_8 \cos\theta_{22} p_9 \\
& \quad \quad - 2 \cos\theta_{22} \cos\theta_8 p_{11} - 2 \cos\theta_{22} \cos\theta_8 p_8 - 2 \cos\theta_{22} \cos\theta_8 p_2, \\
R_{12}(\vec{p}) & = \cos\theta_{10} \cos\theta_9 - 2 \cos\theta_9 \cos\theta_{10} p_2 - 2 \cos\theta_{10} \cos\theta_9 p_8 \\
& \quad \quad  - 2 \cos\theta_9 p_{11} - 2 \cos\theta_{10} p_{14}, \\
R_{13}(\vec{p}) & = \cos\theta_{11} \cos\theta_{23} - 2 \cos\theta_{11} \cos\theta_{23} p_1 - 2 \cos\theta_{23} \cos\theta_{11} p_2 \\
& \quad \quad  + 2 \cos\theta_{11} \cos\theta_{23} p_3 - 2 \cos\theta_{11} \cos\theta_{23} p_{13} \\
& \quad \quad  - 2 \cos\theta_{23} \cos\theta_{11} p_{14} + 2 \cos\theta_{11} \cos\theta_{23} p_{15}, \\
R_{14}(\vec{p}) & = \cos\theta_{12} \cos\theta_{24} - 2 \cos\theta_{12} \cos\theta_{24} p_3 - 2 \cos\theta_{24} \cos\theta_{12} p_4 \\
& \quad \quad - 2 \cos\theta_{24} \cos\theta_{12} p_5 - 2 \cos\theta_{12} \cos\theta_{24} p_6 \\
& \quad \quad - 2 \cos\theta_{24} \cos\theta_{12} p_7 - 2 \cos\theta_{12} \cos\theta_{24} p_8 \\
& \quad \quad - 2 \cos\theta_{12} \cos\theta_{24} p_9 - 2 \cos\theta_{12} \cos\theta_{24} p_{11} \\
& \quad \quad + 2 \cos\theta_{24} \cos\theta_{12} p_{10} - 4 \cos\theta_{12} \cos\theta_{24} p_{12} \\
& \quad \quad - 2 \cos\theta_{12} \cos\theta_{24} p_{15}, \\
R_{15}(\vec{p}) & = \cos\theta_{13} \cos\theta_{14} \cos\theta_{25} - 2 \cos\theta_{25} \cos\theta_{13} p_7 - 2 \cos\theta_{14} \cos\theta_{25} \cos\theta_{13} p_4 \\
& \quad \quad - 2 \cos\theta_{25} \cos\theta_{14} p_{14} - 2 \cos\theta_{13} \cos\theta_{25} \cos\theta_{14} p_2 \\
& \quad \quad - 2 \cos\theta_{14} \cos\theta_{25} p_{13} - 2 \cos\theta_{13} \cos\theta_{14} \cos\theta_{25} p_1 \\
& \quad \quad - 2 \cos\theta_{13} \cos\theta_{25} p_6 + 2 \cos\theta_{14} \cos\theta_{25} p_{15} \\
& \quad \quad + 2 \cos\theta_{13} \cos\theta_{14} \cos\theta_{25} p_3.  
\end{aligned}
\end{equation*}

\section{Optimal angles}
Sec.~\ref{s:optimizing} has been devoted to optimize the calibration of a $\tilde{\cnot}$ via GSC with sequences in Tab.~\ref{t:new_sequences25} considering perfect and imperfect measurements. Here we want to present the concrete values of $\theta_1,\dots,\theta_{25}$ that achieve the minimal mean squared distance in both cases.

With the optimization process presented in Sec.~\ref{s:optimizing}, we have obtained a minimal squared distance in the case of perfect measurements (Eq.~(\ref{e:D2})) of $\D\approx 3.4/N$ with the angles

$$
\begin{array}{ccccccc}
\theta_1/\pi & = & 1.3864, & ~~ & \theta_2/\pi & = & 0.3743, \\
\theta_3/\pi & = & 1.3779, & ~~ & \theta_4/\pi & = & 1.3779, \\
\theta_5/\pi & = & 1.3779, & ~~ & \theta_6/\pi & = & 1.5722, \\
\theta_7/\pi & = & 1.0000, & ~~ & \theta_8/\pi & = & 1.3779, \\
\theta_9/\pi & = & 0.4003, & ~~ & \theta_{10}/\pi & = & 1.0000, \\
\theta_{11}/\pi & = & 0.5000, & ~~ & \theta_{12}/\pi & = & 1.5000, \\
\theta_{13}/\pi & = & 1.0000, & ~~ & \theta_{14}/\pi & = & 0.5000, \\
\theta_{15}/\pi & = & 1.0310, & ~~ & \theta_{16}/\pi & = & 0.2618, \\
\theta_{17}/\pi & = & 0.3087, & ~~ & \theta_{18}/\pi & = & 0.8718, \\
\theta_{19}/\pi & = & 0.5000, & ~~ & \theta_{20}/\pi & = & 0.5000, \\
\theta_{21}/\pi & = & 0.5000, & ~~ & \theta_{22}/\pi & = & 1.5000, \\
\theta_{23}/\pi & = & 0.5000, & ~~ & \theta_{24}/\pi & = & 0.5000, \\
\theta_{25}/\pi & = & 1.5000. & ~~ & & &
\end{array}
$$

Using the same optimization process, but the mean squared distance in Eq.~(\ref{e:D2tilde}), which considers imperfect measurements, we have obtained a minimum at $\tilde{\D}\approx3.6/N$ with the angles

$$
\begin{array}{ccccccc}
\theta_1/\pi & = & 0.4459, & ~~ & \theta_2/\pi & = & 0.4459, \\
\theta_3/\pi & = & 1.3779, & ~~ & \theta_4/\pi & = & 1.3779, \\
\theta_5/\pi & = & 0.6221, & ~~ & \theta_6/\pi & = & 0.5541, \\
\theta_7/\pi & = & 1.0000, & ~~ & \theta_8/\pi & = & 0.6221, \\
\theta_9/\pi & = & 0.5541, & ~~ & \theta_{10}/\pi & = & 1.0000, \\
\theta_{11}/\pi & = & 1.5000, & ~~ & \theta_{12}/\pi & = & 0.5000, \\
\theta_{13}/\pi & = & 1.0000, & ~~ & \theta_{14}/\pi & = & 1.5000, \\
\theta_{15}/\pi & = & 1.5541, & ~~ & \theta_{16}/\pi & = & 0.4459, \\
\theta_{17}/\pi & = & 1.5541, & ~~ & \theta_{18}/\pi & = & 0.4459, \\
\theta_{19}/\pi & = & 1.4993, & ~~ & \theta_{20}/\pi & = & 0.5007, \\
\theta_{21}/\pi & = & 0.5007, & ~~ & \theta_{22}/\pi & = & 1.4993, \\
\theta_{23}/\pi & = & 1.5000, & ~~ & \theta_{24}/\pi & = & 0.5000, \\
\theta_{25}/\pi & = & 1.5000. & ~~ & & &
\end{array}
$$
\vspace{5cm}

\FloatBarrier
\section{GSC Simulation and Single Qubit Gates}
\begin{figure}[h!]
\centering
\includegraphics[scale=1]{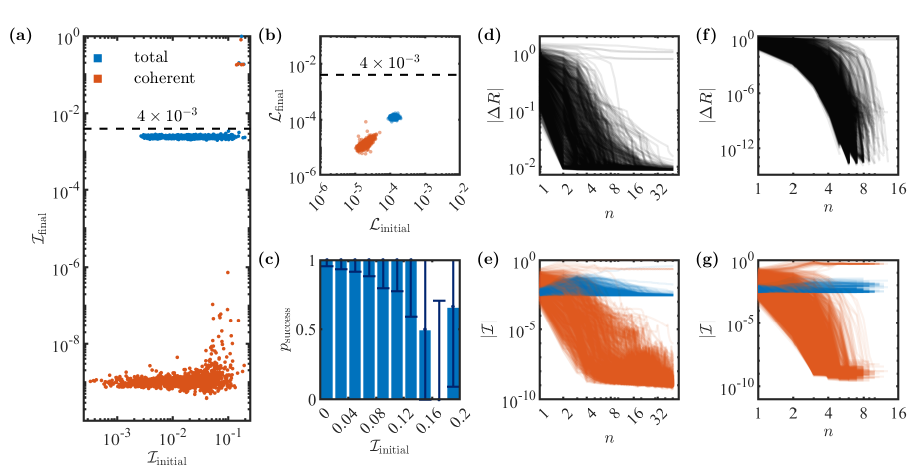}
\caption{Simulation results of optimized GSC in full analogy to the original paper.}
\label{f:result_full}
\end{figure}

\begin{table}[h!]
\centering
\renewcommand\arraystretch{1.25}
\begin{tabular}{ccc}
& $X^{(1)}_{\theta} (T = \SI{20}{ns})$ & $Y^{(1)}_{\theta} (T = \SI{20}{ns})$ \\
\hline\hline
$\alpha$ & 0.7 & 0.7 \\
\hline \hline
$\mathcal{I}_\mathrm{s}$ & 3.79e-04 & 3.32e-04 \\
$\mathcal{I}_\mathrm{f}$ & 6.99e-05 & 5.83e-05 \\
$\mathcal{I}_\mathrm{b}$ & 2.18e-04 & 1.38e-04 \\
$\mathcal{I}_\mathrm{u}$ & 9.31e-13 & 6.60e-13 \\ \hline
$\mathcal{I}_\mathrm{tot}$ & 6.76e-04 & 5.56e-04 \\ \hline\hline
$\mathcal{L}_\mathrm{i}$ & 3.01e-15 & 3.59e-15 \\
$\mathcal{L}_\mathrm{c}$ & 2.22e-15 & 4.44e-16 \\ \hline
$\mathcal{L}_\mathrm{tot}$ & 5.23e-15 & 3.15e-15 \\
\hline\hline
\end{tabular}
\caption{Monte Carlo analysis of the $X^{(1)}_{\theta}$ and $Y^{(1)}_{\theta}$ gates. The columns list the infidelity and leakage contributions for the spectral noise density $S_{\epsilon,\alpha}(f) \propto 1/f^\alpha$, with the experimental value $S_{\epsilon,\alpha}(\SI{1}{MHz}) = \SI{4e-20}{V^2/Hz}$. The figures are calculated using 1000 Monte Carlo time traces with $\SI{3}{\%}$ relative error. The total fidelities calculated with all noise sources applied simultaneously are (from left to right) \SI{99.93}{\%} and \SI{99.94}{\%}. Note that $\mathcal{I}$ includes interaction effects between the noise sources, and is thus not equal to the sum of the first 4 rows. Leakage is much smaller than the infidelity mainly since there is virtually no coupling between the qubits since $b_{23} \gg J_{23}=0$.}
\label{t:best_xtheta_ec0}
\end{table}
\FloatBarrier

\vspace{5cm}
\nocite{*}
\bibliography{bib}
\bibliographystyle{plain}
\end{document}